\documentstyle[12pt]{article}
\begin{document}
\null\vskip-24pt
\hfill SPIN-2000/22
\vskip-01pt
\hfill {\tt gr-qc/0009103}
\vskip0.3truecm
\begin{center}
\vskip 1.5truecm
{\Large\bf
Boundary conditions and the entropy bound
}\\ 
\vskip 1.0truecm
{\large\bf
Sergey N.~Solodukhin\footnote{
email:{\tt S.Solodukhin@phys.uu.nl}}
}\\
\vskip 0.7truecm
{\it Spinoza Institute, University of Utrecht,\\
Leuvenlaan 4, 3584 CE Utrecht, The Netherlands}
\vskip 0.7truemm
\end{center}
\vskip 0.7truecm
\begin{abstract}
\noindent
The entropy-to-energy bound 
is examined for a quantum scalar field confined to a cavity and satisfying Robin  condition
on the boundary of the cavity. It is found that near certain points in the  space of the parameter
defining the boundary condition the lowest eigenfrequency (while non-zero) 
becomes arbitrarily small. Estimating, following Bekenstein and Schiffer,
the ratio $S/E$ by the $\zeta$-function,
$(24\zeta (4))^{1/4}$, we compute $\zeta (4)$ explicitly and find 
that it is  not bounded near those points
that signals  violation of the bound. We interpret our results as imposing certain constraints
on the value of the boundary interaction and estimate the forbidden region in the parameter space
of the boundary conditions.
\end{abstract}
\begin{center}
\end{center}
\vskip 1cm
\newpage
\baselineskip=.81cm

\section{Introduction}

Some time ago Bekenstein proposed \cite{B1}, \cite{B2}
that for a quantum system confined to a cavity of finite size $R$
the entropy to the energy ratio $S/E$ can not be arbitrarily large so that the bound
\begin{equation}
S/E\leq 2\pi R
\label{0.1}
\end{equation}
takes place. Originally the bound was deduced by considering  a {\it gedanken} experiment of
lowering the system into a black hole and demanding this process to satisfy
the generalised second law.
The bound (\ref{0.1}) comes out as a consistency condition between the black hole thermodynamics
and ordinary statistical physics. 
Since the system  initially can be placed far away from the black hole where
the gravitational field is negligible, the bound must hold for any 
system in flat space-time
and be provable with no recourse to gravitational physics.
Thus, the universality of (\ref{0.1}) was conjectured.

The black hole way of deriving the bound was criticised in \cite{UW} (see, however \cite{Bek}).
Nevertheless,
the bound (\ref{0.1}), as it stands, has passed a
number of tests \cite{B1}, \cite{B2}, \cite{SB}, \cite{B3}, \cite{Bx} (for a review see \cite{SB2})
so that its universality deserves the further examination (for a recent discussion see \cite{Page}
and \cite{Bek1}).
In order to make the statement on the entropy bound precise one has to define 
the meaning of $S$, $E$ and $R$
in (\ref{0.1}) as well as the conditions under which the statistical
properties of the system should be considered.
In ref.\cite{SB} the bound (\ref{0.1}) is regarded as applying only to the field in the cavity and it
is  proposed to use the microcanonical methods. One  interprets $S$ as
logarithm of $\Omega (E)$,  the number of quantum states accessible to the field system
with energy up to and including $E$, and ignores the walls of the cavity.

The bound (\ref{0.1}) can  be obviously b exceeded if there are one-particle states with zero-energy
(zero-modes) \cite{Unruh}. Then, by adding arbitrary number of such states one does not change the
total energy of the system but  makes the entropy $S$ arbitrary large. 
The important observation \cite{SB1}, however, is that the zero-mode with some occupation
corresponds to a condensate. The systems with different configurations of the condensate should be
considered as macroscopically different. Thus, only the
excitations with energy above the vacuum should be taken into account, i.e. the zero-modes are to be excluded.

Suppose, the cavity confining the system is circumscribed by a sphere of radius $R$. 
Then, it was shown in \cite{SB} that the microcanonical entropy $S(E)=\ln \Omega (E)$
obeys 
\begin{equation}
S(E)/E<[24 \zeta_{sp}(4)]^{1/4}~~,
\label{0.2}
\end{equation}
where $\zeta_{sp}(k)$ is the $\zeta$-function
$$
\zeta (k)=\sum_{i}\omega_i^{-k}
$$
for the sphere, where $\{\omega_i \}$ is the discrete one-particle energy
spectrum with zero-modes excluded.
Since for the sphere we have $\zeta (4)\sim R^4$ the bound
(\ref{0.1}) follows from (\ref{0.2}) provided $R^{-4}\zeta (4)$ is
appropriately bounded from above. The later was verified in \cite{SB} for various types of
free fields
satisfying Dirichlet or Neumann conditions on the sphere.

In this paper we make a step further and impose more general, of Robin type, condition on the
field on the boundary of the cavity. Note, that
the boundary condition of this type should be always imposed on a quantum field non-minimally coupled to 
the metric.
The simplest case is the scalar field described by the action
$$
W=-{1\over 2}\int_M\left( (\nabla\phi )^2+\xi \phi^2 R\right)-\int_{\partial M}\xi \phi^2 K~~,
$$
where $R$ is Ricci scalar and $K$ is the extrinsic curvature of the boundary $\partial M$.
The boundary term is necessary to add in order to
 the stress-tensor for the theory be well-defined.
Variation of this action 
with respect to $\phi$ gives us not only the equation of motion in the interior
$$
-\nabla^2\phi+\xi R \phi=0
$$
but also the boundary condition
\begin{equation}
\left(n^\mu\partial_\mu\phi+2\xi K\phi\right)_{\partial M}=0
\label{0.3}
\end{equation}
of the Robin type. Note also that the allowing for a more general boundary condition
is in accord with the general assumption of the Schiffer-Bekenstein paper
\cite{SB} that all interactions of the field are negligible ``except for those which confine it and are
expressed as boundary conditions''. The boundary condition (\ref{0.3}) encodes in a generic form such boundary
interaction \cite{BS}.

The $\zeta$-function is simpler object for computation (the computational technique appropriate to the case under consideration
was developed in \cite{Moss} and\cite{Dowker}) 
than the ratio $S(E)/E$. Therefore, in this paper we mainly analyse 
the $\zeta$-function. 
Considering  $R^{-4}\zeta (4)$ as function of the parameter in the Robin boundary condition we
find that there are special points in the parameter space near which this function is
unbounded from above. This can be easily understood. Exactly at those special points
the quantum field has a zero-mode.
When any of these points is approached in the 
parameter space it signals in that the lowest eigenfrequency $\omega_1$, while non-zero, becomes arbitrarily 
small. Since $\omega_1$ makes the dominant contribution to the $\zeta$-function (see also \cite{B2}),  one has
$\zeta (4)\simeq {1\over \omega_1^4}$ if the lowest energy state is non-degenerate. It 
 is evident that $\zeta (4)$ is unbounded in this case.

The same is also true  for the ratio
$S(E)/E$ itself. When the parameter in the boundary condition approaches one of those special points
it is typical 
that in the spectrum appears a large gap between
the lowest (non-zero) eigenfrequency $\omega_1$ and the next 
eigenfrequency $\omega_2$, $\ \omega_2/\omega_1>>1$. When the energy of the system is $E=nE_1$ 
for some integer $n$ and  $E<\omega_2$ only the lowest energy 
level is populated. The number of accessible states (assuming, for simplicity, that $g_1=1$ for the degeneracy
of the lowest energy level) is $\Omega (E)=(n+1)$ and we have
$S(E)/E= {1\over \omega_1}n^{-1}\ln (n+1)$. Since $\max (n^{-1}\ln (n+1))=\ln 2\simeq 0.7$ we obtain that \cite{deg}
\begin{equation}
\max (S(E)/E)\simeq {0.7 \over \omega_1}~~.
\label{0.4}
\end{equation}
This  shows, in particular, that relation (\ref{0.2}) should be considered as a good estimate for the maximum
of $S(E)/E$ rather than just giving an upper bound on $S(E)/E$.
It follows that the bound (\ref{0.1}) holds only if  $R\omega_1$
is restricted from below and is violated if $R\omega_1$ can be made arbitrarily small.
The later does occur for the certain values of the parameter in the boundary condition as we show
in this paper.

In the next section we consider in detail the case of $(1+1)$-dimensional massless field for 
which the analysis  of the spectrum and the computation of $\zeta$-function
are especially simple. The $(3+1)$-dimensional field is analysed in section 3 and the
massive field is briefly discussed in section 4. 
The size of ``forbidden region'' in the parameter space is estimated in section 5.
Section 6 contains some concluding remarks.

\bigskip

\section{The interval}

We start our analysis with consideration of $(1+1)$-dimensional massless scalar field
living on the interval $[0,R]$ with the boundary conditions
\begin{equation}
\left( {d\over dx} \phi +{h_0\over R}\phi\right)_{x=0}=0~~,~~~ 
\left( {d\over dx}\phi -{h_1\over R}\phi \right)_{x=R}=0~~.
\label{1}
\end{equation}
The energy eigenfunction $\phi_{\omega}(x)e^{\imath \omega t}$ satisfies
the differential equation
\begin{equation}
-{d^2\over dx^2}\phi_\omega (x)=\omega^2 \phi_\omega (x)
\label{2}
\end{equation}
and takes the form
$$
\phi_\omega (x)=N\sin (\omega x+\delta )~~.
$$
The spectrum is discrete, $\omega_n=\alpha_n R^{-1}$, where $\{ \alpha \}$ are (positive) 
roots of the equation
\begin{equation}
{\tan{\alpha}\over \alpha}={h_0+h_1\over h_0h_1-\alpha^2}~~.
\label{3}
\end{equation}
In general, there also may be  bound states (for which $\omega^2=-\lambda^2 R^{-2}<0$)) with wave function
$\phi_\lambda (x)=Ae^{{\lambda\over R}x}+Be^{-{\lambda\over R}x}$. The equation on $\lambda$ is
\begin{equation}
 {\tanh{\lambda}\over \lambda}={h_0+h_1\over \lambda^2-h_0h_1}~~.
\label{4}
\end{equation}

As a consequence of the Mittag-Leffler theorem, we have \cite{Dowker}
\begin{equation}
(h_0h_1-z^2){\sin{z}\over z}-(h_0+h_1)\cos{z}=(h_0h_1-h_0-h_1)\prod_{\alpha >0} (1-{z^2\over \alpha^2})~~.
\label{5}
\end{equation}
This formula helps to evaluate explicitly the sums of inverse powers of the roots. One just has 
to take the logarithm of both sides of eq.(\ref{5}), expand in powers of $z^2$ and equate the relevant 
coefficients. The expressions for any $h_0$ and $h_1$ are given in \cite{Dowker}. To simplify the things,
in what follows,  we assume that $h_0=0, h_1=h$.
One finds,
\begin{equation}
\sum_\alpha {1\over \alpha^2}={1\over 2}-{1\over h}~~,
\label{6}
\end{equation}
\begin{equation}
\sum_\alpha {1\over \alpha^4}={1\over h^2}-{2\over 3h}+{1\over 6}~~.
\label{7}
\end{equation}
Let us consider the case of positive $h$ first.
For $h_0=0, h_1=h$ the equation (\ref{4}) becomes
\begin{equation}
\lambda\tanh\lambda=h
\label{8}
\end{equation}
and we see that for positive $h$ there is one bound state \cite{comment1}, $\lambda=\lambda_b$.
When $h$ is close to zero its value  expands in powers of $h$
$$
\lambda^2_b=h+{1\over 3}h^2+{4\over 45}h^3+O(h^4)
$$
and hence one has
\begin{eqnarray}
&&-{1\over \lambda^2_b}=-{1\over h}+{1\over 3}+O(h)~~,\nonumber \\
&&{1\over \lambda^4_b}={1\over h^2}-{2\over 3h}+{7\over 45}+O(h)~~~.
\label{9}
\end{eqnarray}
It is important to note that  for positive $h$ expressions (\ref{6}) and (\ref{7}) obtained by using the 
Mittag-Leffler theorem
include the contribution of the bound state \cite{comment}. 
The small $h$ behaviour of (\ref{6})
and (\ref{7}) is due to the bound state as one can see by comparing (\ref{6}), (\ref{7}) and (\ref{9}).
In particular, this explains why the expression (\ref{6}) is negative when $0<h<2$.

The $\zeta$-function we want to compute is defined for the part of the spectrum with $\omega^2>0$.
Hence we have to exclude the bound state.  Subtracting ${1\over \lambda^4_b}$ from (\ref{7})
we get
\begin{equation}
R^{-4}\zeta (4)\equiv \sum_{\alpha^2>0}{1\over \alpha^4}={1\over 80}+{4\over 945}h-{1\over 675}h^2+O(h^3)~~.
\label{10}
\end{equation}
The divergence of the  sum
 (\ref{7}) at small $h$ is due to the bound state. After 
the subtraction  the sum becomes finite at $h=+0$. In fact, it is  bounded
for all $h\geq 0$, monotonically increasing from ${1\over 80}$ at small $h$ to ${1\over 6}$ 
for infinitely large $h$.
We conclude that for positive $h$ the function $R^{-4}\zeta (4)<{1\over 6}$
and the Bekenstein bound perfectly holds.

Consider now the case of negative $h$. There is no bound state in this case, so that the expression
(\ref{7}) gives us exactly  $R^{-4}\zeta (4)$. As function of $h$ it approaches
${1\over 6}$  at $h\rightarrow -\infty$  and grows as $({1\over h^2}-{2\over 3h})$
when $h$ is close to $-0$.
Thus, there is no upper bound for $\zeta (4)$ and the universal entropy bound 
can not hold in this case.

It is easy to understand why this happens. The lowest root of the equation (\ref{3}), which for $h_0=0$, 
$h_1=h$ reads
\begin{equation}
\alpha \tan\alpha=-h~~,
\label{11}
\end{equation}
is always between $0$ and $\pi /2$ when $h<0$. For small negative $h$ 
\begin{equation}
\alpha^2_1=-h-{1\over 3}h^2-{4\over 45}h^3+O(h^4)
\label{12}
\end{equation}
it approaches zero and when $h=0$ it becomes the known zero-mode of the Neumann boundary
value problem. A higher eigenvalue $\alpha_n,~n>1$ lies between ${\pi\over 2}(n-1)$ and ${\pi\over 2}n$
for all values of $h$. The value of $\zeta (4)$ for small $h$ is then mostly due to $\alpha_1$.
Indeed, approximating
$$
R^{-4}\zeta (4)\simeq \alpha_1^{-4}={1\over h^2}-{2\over 3h}+{7\over 45}+O(h)
$$
we find agreement with (\ref{7}) up to
$O(h^0)$ terms.  
The $\zeta (4)$  is unbounded  
(that indicates a violation of the entropy-to-energy  bound) because the lowest energy level 
$\omega_1=\alpha_1R^{-1}$
of the spectrum can be made arbitrarily small.

For positive $h$ there is no state with energy close to zero, the lowest $\alpha$
appearing in the interval between $\pi /2$ and $\pi$. 
When $h$ passes from the  negative to positive values the lowest excited one-particle
state (with eigenfrequency $\omega_1=\alpha_1R^{-1}$)
becomes the bound state (with $\omega_b^2=-\lambda^2_bR^{-2}$) and should be excluded. 
For positive $h$, this saves the entropy 
bound.
However, the possibility to make the lowest energy level arbitrarily small 
is fatal for the validity of the bound  when $h$ is negative.

\bigskip

\section{The 3D ball}

In flat $(3+1)$-dimensional space-time consider a massless scalar field confined to a spherical
cavity of radius $R$. The boundary condition in this case is
\begin{equation}
\left( {d\over dr}\phi-{h\over R}\phi \right)_{r=R}=0~~,
\label{2.1}
\end{equation}
where $r$ is the radial coordinate. 
The condition (\ref{2.1}) has the form (\ref{0.3}) with $\xi=-{1\over 4}h$ since for the sphere the extrinsic curvature is
$K={2\over R}$.The Dirichlet boundary condition corresponds to infinite $h$.
The energy 
eigenfunction $\phi_\omega=f_\omega (r)Y_{l,n}(\theta , \varphi )e^{\imath \omega t}$ 
expands in terms of the spherical harmonics $Y_{l,n}(\theta ,\varphi )$,
the degeneracy being $(2l+1)$. The equation on the radial function reads
\begin{equation}
{1\over r^2}\partial_r (r^2\partial_r f_\omega )-{l(l+1)\over r^2}f_\omega =-\omega^2f_\omega~~.
\label{2.2}
\end{equation}
Solutions to this equation should satisfy the boundary condition (\ref{2.1}) and be regular at
$r=0$. There are three types of such solutions.

{\bf 1. Zero-modes}

When $\omega=0$ two solutions are possible, $f_1(r)=r^l$ and $f_2(r)=r^{-l-1}$, only the first being
regular at $r=0$. The function $f_1(r)$ satisfies the boundary condition (\ref{2.1}) only if
the parameter $h$ in  (\ref{2.1}) is some (non-negative) integer,  $h=l_0$. Then the zero mode exists for $l=l_0$
and  $f_1(r)=r^{l_0}$. For $h=0$ this solution is the known zero-mode of the Neumann
problem. The similar zero-modes appear every time when $h$ is a positive integer $l_0$.
In total, there are $(2l_0+1)$ of them.

{\bf 2. Propagating modes}

For $\omega^2>0$ the solution regular at $r=0$ is
$$
f_\omega (r)=N_\omega r^{-1/2}J_{l+{1\over 2}}(\omega r)~~,
$$
where $N_\omega$ is normalisation constant. Since this solution should satisfy the boundary condition
(\ref{2.1}) the spectrum is discrete $\omega_{n,l}=\alpha_{n,l}R^{-1}$, where
$\{\alpha_{n,l}\}$ are the roots of the equation 
\begin{equation}
h=l-\alpha {J_{l+{3\over 2}}(\alpha )\over J_{l+{1\over 2}}(\alpha )}~~.
\label{2.4}
\end{equation}
We have learned from the $(1+1)$-dimensional example considered above that 
we have to watch for the energy level which may be arbitrary close
to zero. For a fixed $l$ the point $\alpha=0$ is the point where the function staying 
at the right hand side of eq.(\ref{2.4})
takes the maximal value equal $l$. For negative $h$ the lowest root which may be close to zero is
$\alpha_{1,0}$ corresponding to $l=0$. One has $\alpha_{1,l}>2$ for  $l\geq 1$. 
When $h$ becomes positive but less
than $1$ such root appears at $l=1$, $\alpha_{1,1}$. In general, for positive
$h$ lying in the interval $l_0-1<h<l_0$, where $l_0$ is positive integer, the lowest root is
$\alpha_{1,l_0}$ corresponding to $l=l_0$.
For small $(l_0-h)>0$ one finds 
\begin{equation}
\alpha^2_{1,l_0}=2(l_0+{3\over 2})(l_0-h)-{(l_0+{3\over 2})\over (l_0+{5\over 2})}(l_0-h)^2
+{(l_0+{3\over 2})\over (l_0+{5\over 2})^2(l_0+{7\over 2})}(l_0-h)^3
+O(l_0-h)^4~~.
\label{*}
\end{equation}

{\bf 3. Bound states}

These are the regular at $r=0$ solutions with $\omega^2<0$,
\begin{equation}
f_\lambda (r)=N_\lambda r^{-1/2}I_{l+{1\over 2}}({\lambda \over R}r)~~,
\label{f}
\end{equation}
where $\lambda$ should be  determined from equation
\begin{equation}
h=l+\lambda {I_{l+{3\over 2}}(\lambda )\over I_{l+{1\over 2}}(\lambda )}~~.
\label{2.6}
\end{equation}
The function of $\lambda$ staying at the right hand side of eq.(\ref{2.6}) takes at $\lambda=0$
its minimal value equal $l$
and grows monotonically as linear function 
for large $\lambda$. The solution to the equation (\ref{2.6}), thus, exists
only when $h$ is positive and for a given $l$ there may be no more than one such solution.
If $h$ lies in the interval $l_0\leq h<l_0+1$, where $l_0$ is non-negative integer, solution to the equation
(\ref{2.6}) exists for $l=0,...,l_0$. The total number of the bound states (taking into account
the degeneracy due to angles) is $\sum_{l=0}^{l_0}(2l+1)=(l_0+1)^2$.
For small $(h-l_0)>0$ one has
\begin{equation}
\lambda^2_{l_0}=2(l_0+{3\over 2})(h-l_0)+{(l_0+{3\over 2})\over (l_0+{5\over 2})}(h-l_0)^2
+{(l_0+{3\over 2})\over (l_0+{5\over 2})^2(l+{7\over 2})}(h-l_0)^3
+O(h-l_0)^4~~.
\label{*1}
\end{equation}
We see from eqs.(\ref{*}) and (\ref{*1}) that when $h$ passes through the point $h=l_0$ the 
propagating state with the
lowest
eigenfrequency  $\omega_1^2=R^{-2}\alpha^2_{1,l_0}$  becomes a bound state 
with $\omega_b^2=-R^{-2}\lambda^2_{l_0}$.

\bigskip

Evaluating sums of inverse powers of the roots we can again employ the Mittag-Leffler theorem.
At  fixed $l$ one has
\begin{equation}
z^{-(l+{1\over 2})}F_{l+{1\over 2}}(z)=\gamma_l \prod_{\alpha >0}(1-{z^2\over \alpha^2})~~,
\label{2.7}
\end{equation}
where $\gamma_l$ is some constant (see \cite{Dowker}) and $\{ \alpha \}$ are the roots (\ref{2.4}),
$F_{l+{1\over 2}}(\alpha )=0$, where
\begin{equation}
F_{l+{1\over 2}}(z)=zJ'_{l+{1\over 2}}(z)+(l-h)J_{l+{1\over 2}}(z)~~.
\label{2.8}
\end{equation}
Up to terms of order $z^{l+{1\over 2}+6}$ it  expands as follows
\begin{eqnarray}
F_{l+{1\over 2}}(z)={z^{l+{1\over 2}}\over 2^{l+{1\over 2}}\Gamma ({l+{3\over 2}})} \left(
1-{1\over 4(l+{3\over 2})}{(l-h+2)\over (l-h)}z^2+
{1\over 32 ({l+{3\over 2}})({l+{5\over 2}})}
{(l-h+4)\over (l-h)}z^4 \right)
\label{2.9}
\end{eqnarray}
Taking the logarithm of (\ref{2.7}), expanding  in powers of $z^2$ and using (\ref{2.9})
one gets
\begin{equation}
\sum_{n=1}^\infty{1\over \alpha^4_{n,l}}={1\over 16}\left( {1\over ({l+{3\over 2}})^2}
{(l-h+2)^2\over (l-h)^2}-{1\over ({l+{3\over 2}})({l+{5\over 2}})}{(l-h+4)\over (l-h)} \right)
\label{2.10}
\end{equation}
for the sum  at fixed $l$. In order to evaluate the $\zeta$-function
we have to sum over all possible $l$ and take into account the degeneracy
\begin{equation}
R^{-4}\zeta (4)\equiv \sum_{l=0}^\infty (2l+1)\sum_{  n } ~~{1\over \alpha^4_{n,l}}~~.
\label{2.11}
\end{equation}
We remind that the $\zeta$-function is defined for the spectrum with $\alpha^2>0$.
The zero-modes and bound states are, thus, to be  excluded in (\ref{2.11}).

When $h$ is negative there are no bound states or zero-modes. Substituting (\ref{2.10}) into
(\ref{2.11}) the sum over $l$ can be computed  explicitly and is derived in terms of  psi-function
as follows
\begin{eqnarray}
&&R^{-4}\zeta (4)={2\over 3}{(2h+1)(4h^2-5)\over (2h+3)^3}+{(2h+1)\over (2h+3)^2}\Psi (1,-h)-\nonumber \\
&&{16\over (2h+5)(2h+3)^3}(\Psi (-h)-{8\over 3}+\gamma +2\ln 2)-{\pi^2\over 16}
{(2h-1)^2\over (2h+3)^2}~~.
\label{S}
\end{eqnarray}
It seems  that the function (\ref{S}) has a pole at $h=-3/2$ and $h=-5/2$. However, it is easy to check that
it is regular at those points. The function (\ref{S}) monotonically increases from the
value ${2\over 3}-{\pi^2\over 16}$ at $h=-\infty$ (Dirichlet boundary condition) to
infinity when $h$ approaches zero. We conclude that the zeta-function is unbounded near $h=0$.
The expression (\ref{S}) is valid also for the positive $h$. However, in this case
there may be bound states and zero-modes (their appearance signals in that (\ref{S}) has poles
at integer $h$)
the contribution of which must be subtracted from  the right hand side of 
(\ref{S}).

When $h$ approaches from below  any non-negative integer $l_0$ the expression (\ref{2.10}) for $l=l_0$
grows to infinity. It is due to the fact that the lowest root (\ref{*}), $\alpha_{1,l_0}$,
becomes arbitrarily small. Indeed, approximating the sum (\ref{2.10}) for $l=l_0$
 by $1/\alpha^4_{1,l_0}$
we find that
\begin{equation}
\sum^\infty_{n=1}{1\over \alpha^4_{n,l_0}}-{1\over \alpha^4_{1,l_0}}=
{1\over 16 (l_0+{5\over 2})^2(l_0+{7\over 2})}+O(h-l_0)~~.
\label{2.12}
\end{equation}
Thus, for $h$ less but very close to $l_0$ one finds that the $\zeta$-function 
\begin{equation}
R^{-4}\zeta (4)={(2l_0+1)\over \alpha^4_{1,l_0}(h)}+O(l_0-h)^0~~,
\label{2.13}
\end{equation}
where $\alpha^2_{1,l_0}$ is given by eq.(\ref{*}),
is not bounded from above. This also can be seen from the analysis of the poles in (\ref{S}).

The whole picture changes when $h$ passes through the point $h=l_0$ and becomes  slightly grater 
than $l_0$. In this case the root $\alpha_{1,l_0}$ disappears. But, instead, there
appears a bound state, $\lambda^2_{l_0}$. 
In this case, as  was explained in the previous section,
 the expression (\ref{2.10}) for $l= l_0$  (and, hence, also the right hand side of eq.(\ref{S}))
contains the contribution of the 
bound state which  should be 
excluded \cite{note}.
The sum $\sum_n 1/\alpha^{4}_{n,l_0}$ then is over the propagating modes only and is finite
when $h$ approaches $l_0$ from above. As a result, the $\zeta$-function (\ref{2.11}) is  finite when
$h\rightarrow l_0+0$.
A special case is when $h=l_0$ exactly. In this case there appears a zero-mode $f_1=r^{l_0}$ which should be
excluded when one computes the $\zeta$-function. 
When $h$ increases, this zero-mode becomes the bound state corresponding to $l=l_0$ while 
the rest of the spectrum
changes continuously. Therefore, the case $h=l_0$ can be achieved by taking the limit $h\rightarrow l_0+0$.
All this repeats every time when $h$ becomes close to some positive integer.
The general behaviour of the spectrum and the $\zeta$-function near every such point is similar to that 
we had in $(1+1)$-dimensional case.
We conclude that in a way similar to the  two-dimensional case the four-dimensional  $\zeta (4)$ 
considered as function of the parameter $h$ in the boundary condition (\ref{2.1}) is not bounded
near each point $h=l_0$ at which the radial equation (\ref{2.2}) has a zero-mode.

\bigskip

\section{The massive field}

In this section we briefly discuss the massive field. In the presence of mass $m=\mu R^{-1}$
the eigenfrequencies $\omega_\alpha$ are defined as $R^2\omega^2_\alpha=\mu^2+\alpha^2$,
where $\{ \alpha \}$ are the roots considered in section 2. The $\zeta$-function then reads
\begin{equation}
R^{-4}\zeta (4)=\sum_{\{\alpha \}, \ \mu^2+\alpha^2 >0}{1\over ( \mu^2+\alpha^2)^2}~~,
\label{3.1}
\end{equation}
where  sum is over states with $\omega^2_\alpha>0$. We see that due to 
mass some of the bound states of the massless field become now the 
propagating states. In 
particular, the zero-mode  is the state (\ref{f}) with $\lambda=\mu$.
The partial summation in (\ref{3.1}) for fixed $l$  can again be done 
with  the help of the Mittag-Leffler theorem \cite{Dowker}. After analytical continuation
 $z\rightarrow \imath \mu$  eq.(\ref{2.7}) reads
\begin{eqnarray}
&&\mu^{-(l+{1\over 2})}\Phi_{l+{1\over 2}}(\mu )=\gamma_l \prod (1+{\mu^2\over \alpha^2})~~,
\nonumber \\
&&\Phi_{l+{1\over 2}}(\mu )=\mu I'_{l+{1\over 2}}(\mu )+(l-h)I_{l+{1\over 2}}(\mu )~~.
\label{3.2}
\end{eqnarray}
Taking the logarithm of (\ref{3.2}) and differentiating with respect to $\mu^2$ one
finds
\begin{equation}
\sum_{\{\alpha \}, \ fixed \ l}{1\over (\mu^2+\alpha^2 )^2}=-{d\over d\mu^2}
\left({d\over d\mu^2} \ln (\mu^{-(l+{1\over 2})}\Phi_{l+{1\over 2}}(\mu ))\right)~~.
\label{3.3}
\end{equation}
Of course, using (\ref{3.3}) for evaluation the sum (\ref{3.1}) we should watch for the bound states
($\mu^2+\alpha^2<0$) and zero-modes ($\mu^2+\alpha^2=0$) and subtract their contribution from 
the right hand side of 
(\ref{3.3}).  We have two parameters in our disposal, $h$ and $\mu$. When $h$ is negative 
there are no states with $\mu^2+\alpha^2\leq 0$. Let us fix $h$, $l_0\leq h < l_0+1$,
where $l_0$ is some non-negative integer. In this case, there are $(l_0+1)$ roots 
$\lambda_{l_0}<...<\lambda_1 < \lambda_0$ of the eq.(\ref{2.6}),
$$
\Phi_{l+{1\over 2}}(\lambda )=0~~.
$$
A zero-mode of the massive field appears when 
$\mu$ equals to one of these $\lambda$, say $\mu=\lambda_k$.
For $\mu $ close to  $\lambda_k$ one finds that
$$
\Phi_{l+{1\over 2}}(\mu )=C_k(\mu^2-\lambda_k^2)+O(\mu-\lambda_k)^3~~,
$$
where $C_k$ is some constant.

Substituting this expression into (\ref{3.3}) we find that the sum
\begin{equation}
\sum_{\{\alpha \}, \ fixed \ l}{1\over (\mu^2+\alpha^2 )^2}={1\over (\mu^2-\lambda_k^2)^2}
+O(\mu-\lambda_k )^0~~
\label{3.4}
\end{equation}
diverges when $\mu$ approaches $\lambda_k$. If $\mu$ is slightly grater than $\lambda_k$,
the $\lambda_k$-state is a propagating state. In  fact it is the state with the lowest positive
eigenfrequency, $\omega^2_1=\mu^2-\lambda^2_k$. The expression (\ref{3.4}) is divergent   because
the lowest energy level $\omega_1$ can be  arbitrarily small. This results in
 the corresponding
divergence of the zeta-function (\ref{3.1}).

Decrease now $\mu$ so that it becomes slightly less than $\lambda_k$. 
Then the $\lambda_k$-state 
becomes a bound state of the massive field.
Subtracting its contribution from (\ref{3.4}) and (\ref{3.3}) we get a finite
expression. This means that the $\zeta$-function is bounded from above as 
$\mu\rightarrow \lambda_k-0$. This resembles the picture we had in the previous section when
parameter $h$ approached a positive integer. In both cases the $\zeta$-function
is unbounded in half-vicinity of a point in the parameters space where a zero-mode appears.
Also, it  is so because the lowest eigenfrequency $\omega_1$
is arbitrarily small when that point is approached.

\bigskip

\section{The estimate for ``forbidden region''}

It follows from the analysis just given that the bound (\ref{0.1})
can not hold for all values of the parameter $h$ in the boundary condition (\ref{2.1}).
However, since $\zeta (4)$ as function of $h$ is sharply picked at integer values of $h$, one may expect 
that the bound (\ref{0.1}) holds for almost all values of $h$ 
except a narrow region near integer $h$. Then (\ref{0.1})
would be still valid for a ``typical'' boundary condition. It is indeed the case. In order to
show this we need to estimate the ``forbidden region'' in the space of boundary conditions where
the bound is violated. In order to make the estimate as precise as possible it is the best to use the
formula (\ref{0.4}) and its generalisation \cite{deg} for arbitrary degeneracy.
Using (\ref{*}) we find for $h$ close to an integer $n$ that
\begin{equation}
\max (S(E)/E)\simeq R {\ln g_n \over \sqrt{(2n+3)(n-h)}}~~,
\label{5.1}
\end{equation}
where  $g_n=2n+2$. 
The bound (\ref{0.1}) is exceeded for $h$ lying in the interval  $n-\epsilon_n\leq h<n$,
where for $\epsilon_n$ we find using (\ref{5.1}) the following estimate
\begin{equation}
\epsilon_n\simeq {1\over 4\pi^2}{(\ln g_n)^2\over (2n+3)}~~.
\label{5.2}
\end{equation}
In particular, we find that $\epsilon_0\simeq 0.00406$, $\epsilon_{10}\simeq 0.0105$ and 
$\epsilon_{100}\simeq 0.00352$.
We see that the ``forbidden region'' near each integer $n$ is indeed very
narrow and shrinks for large $n$. It follows that the bound (\ref{0.1}) holds for almost all negative
values of the parameter $h$ except the narrow region $-\epsilon_0<h<0$. Recalling the relation
between $h$ and the non-minimal coupling $\xi$ in (\ref{0.3}), $h=-4\xi$,
we find that the forbidden region for positive $\xi$ is from $0$ to $10^{-3}$. Of course, 
the conformal coupling 
$\xi={1\over 6}$ is quite far from this region.

For positive $h$ there is a piece of the forbidden region near each integer value of $h$. 
In order to estimate how
dense this region is  in the parameter space, let us  consider large interval $0<h<N$.
The relevant density then is given by quantity ${1\over N}\sum_{n=1}^N \epsilon_n$.
Using (\ref{5.2}) we find 
\begin{equation}
{1\over N}\sum_{n=1}^N \epsilon_n\simeq {1\over 24\pi^2}{1\over N}(\ln N )^3~~.
\label{5.3}
\end{equation}
for large $N$.
So, picking randomly a boundary condition with a positive $h$ we almost always put finger on
the right one for which the bound (\ref{0.1}) comes ok. 
For example, for the interval from $0$ to $N=100$ the probability to choose a wrong condition is
$4\times 10^{-3}$.
Thus, the bound (\ref{0.1}) still holds for a ``typical'' boundary condition.


\section{Concluding remarks}

Recall the basic assumptions (see \cite{SB}):
i) $S(E)$ is microcanonical entropy defined as logarithm of the number of 
accessible states with energy up to $E$; 
ii) $E$ is the energy over vacuum (zero-modes are excluded); 
iii) all interactions are negligible except the boundary interaction expressed as
a boundary condition; 
iv) walls of the cavity are ignored.
Among these assumptions, the last one is  perhaps  the most suspicious
although we just follow the prescriptions of paper \cite{SB}.
By the original idea of Bekenstein  (recently re-stated in \cite{Bek1})
the bound (\ref{0.1}) applies to a complete system
with $E$ being the total gravitating energy 
of the system. So it might be that when the boundary condition of the form (\ref{0.3}) is imposed
the boundary itself may carry some part of the total energy and, possibly, entropy.  
The bound (\ref{1}) then should apply to the complete system of the quantum field and walls.
This possibility  should be  further investigated.

Alternatively, one may try to redefine the ground state. Consider for simplicity the two-dimensional field system 
analysed in section 2. The lowest energy level can be 
defined as a new vacuum, then the  next excited level has energy
$(\omega_2-\omega_1)$ which is bounded from below for all values of $h$.
So that the entropy bound holds in this case. 
However, the gravitating energy is not something one can define by a random choice of zero.
It is not clear at the moment how  the gravitating energy is actually computed for the system under consideration.
Although, there still might be possible, by changing the rules appropriately, to save the bound for
all boundary conditions we prefer to interpret our results, provided the bound applies to the field only, 
as imposing certain (quite relaxed)
constraints on the value of the boundary interactions.


{\bf Acknowledgements:} I would like to thank J.D. Bekenstein, J.S. Dowker, G. 't Hooft and 
M. Parikh
for useful discussions. 
I thank Bogoliubov Laboratory of Theoretical Physics at JINR, Dubna
for the hospitality while this work was in progress.


\begin{thebibliography}
\bibitem{deg} For arbitrary degeneracy $g_1>1$ we find that $\max (S(E)/E)={\ln (g_1+1)\over \omega_1}$.
\bibitem{comment1} In general, there is one bound state if the following conditions are satisfied:
$-h_0h_1 \geq 0$, $h_0+h_1\geq 0$ and ${h_0+h_1\over -h_0h_1}\geq 1$; two bound states
if $h_0+h_1>0$ and $-h_0h_1 <0$ and no bound states in any other case.
\bibitem{comment} It is particularly easy to see for a special case $h_0+h_1=0$. The bound
state then is $\lambda^2_b=-h_0h_1$ and the roots of eq. (\ref{3}) are $\alpha_n=\pi n, n=1,2,...$.
Including both the bound state and the states with $\alpha^2>0$ 
it is straightforward to evaluate the sum
$-{1\over \lambda^2_b}+\sum_{n=1}^{\infty}{1\over \alpha^2_n}={1\over h_0h_1}+{1\over 6}$. It exactly 
coincides with what we get by expanding the logarithm of eq.(\ref{5}) in powers of $z^2$.
The contribution of the bound state is evident. I thank J.S. Dowker for this example.
\bibitem{note} One should also subtract the contribution of all other bound states appearing for $l=0,..,l_0-1$.
\end{thebibliography}
\end{document}